\documentclass{iitpressproc}

\usepackage{graphicx}
\usepackage{amsmath}

\def\bea{\begin{eqnarray}}
\def\eea{\end{eqnarray}}

\begin{document}

\title{Physics implications of correlation data from the RHIC and LHC heavy-ion programs}

\author{R. L. Ray
\address{Department of Physics, University of Texas at Austin, Austin, TX 78712 USA}\\[2ex]
}

\maketitle

\begin{abstract}
Two-dimensional angular correlation data from the STAR experiment at RHIC and from the LHC experiments provide critical information about dynamical processes in relativistic heavy-ion collisions.  The principal correlation structures of interest are a broad jet-related peak at small relative azimuth ($\phi$) extending to large relative pseudorapidity ($\eta$), the dijet ridge at large relative azimuth, and an independent double ridge on $\phi$ represented by a quadrupole function. The broad peak at small relative azimuth has been attributed, in large part, to initial-state fluctuations and hydrodynamic flow which produce higher-order harmonics on $\phi$. That conjecture is challenged in this paper. It is shown that the net effect of additional higher harmonic model elements is to describe small, non-Gaussian (NG) shapes in the broad jet-related peak.  The quadrupole correlation, which is also conventionally attributed to hydrodynamic flow, is considered within the Balitsky - Fadin - Kuraev - Lipatov (BFKL) Pomeron framework. Preliminary results using this model for the quadrupole correlation for particle production from 200 GeV $p+p$ collisions are shown to be consistent with recent data from STAR.  
\end{abstract}

\section{Introduction}

One of the more interesting observations to emerge from the study of two-particle angular correlation data from heavy-ion collisions at the RHIC and the LHC is the appearance of a two-dimensional (2D) peak at small relative azimuth (same-side $\phi$) which significantly increases in amplitude and in width along relative pseudorapidity for more-central collisions \cite{aya,axialCI,Joern,atlas}. For minimum-bias $p+p$ collisions and for Au+Au collisions from peripheral to mid-centrality (50\% of fractional cross section) at RHIC this correlation peak structure is consistent with perturbative quantum chromodynamics (pQCD) predictions for minimum-bias jets (those with no lower momentum cut) assuming binary nucleon-nucleon collision scaling \cite{axialCI,Tomjetfrag,Tommodfrag}. The dynamical origin of the $\eta$ width increase of the same-side peak for more-central collisions is not known. Alver and Roland \cite{AlverRoland} conjectured that the $\eta$-elongation is caused by triangular flow, a $\cos3(\phi_1 - \phi_2)$ element or sextupole. Critical evaluations~\cite{axialCI,SextBash,Tomv3-1,Tomv3-2} show that this sextupole is determined by the multipole decomposition of the azimuth projection of the same-side 2D peak, implying that the sextupole derives from that structure rather than from some other aspect of the data.

Another long-range $\eta$ correlation is the quadrupole, a $\cos2(\phi_1-\phi_2)$ element, proportional to $v_2^2$ and conventionally attributed to elliptic flow. However, the simultaneous occurrence of pQCD minijets and large quadrupole in peripheral to mid-central 200 GeV Au+Au correlation data~\cite{axialCI} calls into question the notion of a strongly interacting medium.  Analysis of the quadrupole correlation systematics with respect to collision energy, transverse momentum ($p_t$), and centrality shows that its amplitude scales with the number of binary nucleon-nucleon collisions, $\log(\sqrt{s})$ and eccentricity implying that the quadrupole is generated in the initial state rather than via final-state scattering. The properties of the same-side 2D peak and the quadrupole correlation lead to a general consideration of long-range pseudorapidity correlations from heavy-ion collisions and the possibility that these structures can be understood within a pQCD framework.

\section{Analysis method}

For the correlations shown here sibling pairs (those from the same event) and mixed-event pair histograms for all charged particles in the STAR TPC acceptance ($p_t > 0.15$~GeV/$c$, $|\eta| < 1$ and $2\pi$ azimuth) are filled on relative azimuth $\phi_\Delta = \phi_1 - \phi_2$ and pseudorapidity $\eta_\Delta = \eta_1 - \eta_2$. There is no ``trigger'' particle; all pairs are used. A per-particle normalization is used which eliminates the trivial combinatoric $1/N_{ch}$ dependence of per-pair quantities such as $v_2^2$. The measured quantity reported in~\cite{axialCI} is
\bea
\label{Eq1}
\frac{\Delta\rho}{\sqrt{\rho_{\rm ref}}} & \equiv & \sqrt{\rho_{\rm ref}}
\frac{\rho_{\rm sib} - \rho_{\rm mix}}{\rho_{\rm mix}},
\eea
where $\sqrt{\rho_{\rm ref}} = d^2N_{\rm ch}/d\eta d\phi$ is the single charged particle density.

The principal correlation structures are well described with a same-side 2D Gaussian, an away-side ($|\phi_\Delta| > \pi/2$) dipole, a quadrupole, and a same-side 2D exponential which describes conversion electrons and quantum correlations. The standard fitting model is defined in \cite{axialCI,SextBash}. No additional model elements are required to describe the minimum-bias $p_t$-integral 2D angular correlation data from STAR. An added sextupole term would have the form $2A_{\rm S} \cos(3 \phi_\Delta)$.

\section{Same-side 2D peak results}

\begin{figure*}[t]
\centering
\includegraphics[keepaspectratio,width=1.6in]{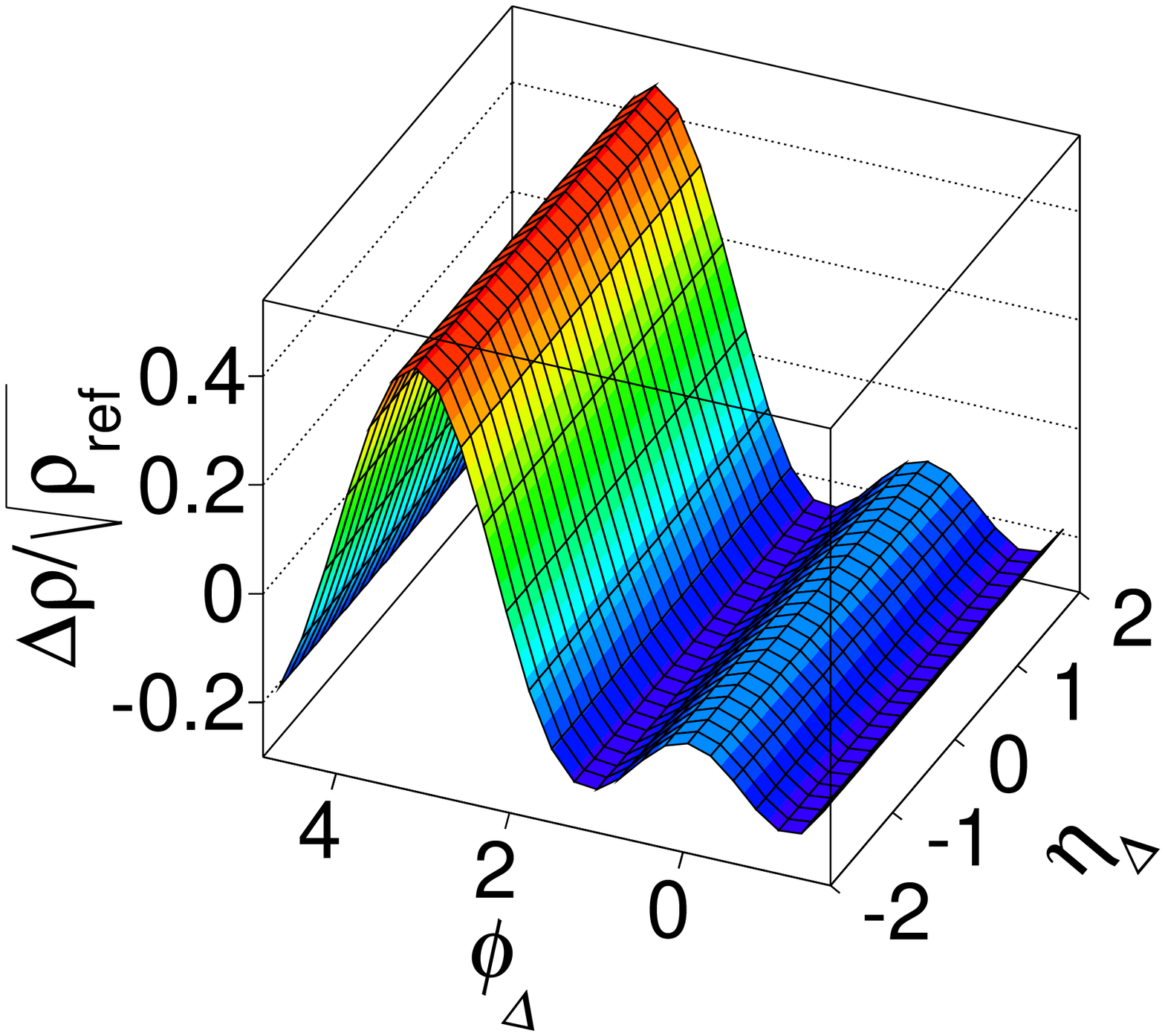}
\put(-40,75){\bf (a)}
\includegraphics[keepaspectratio,width=1.6in]{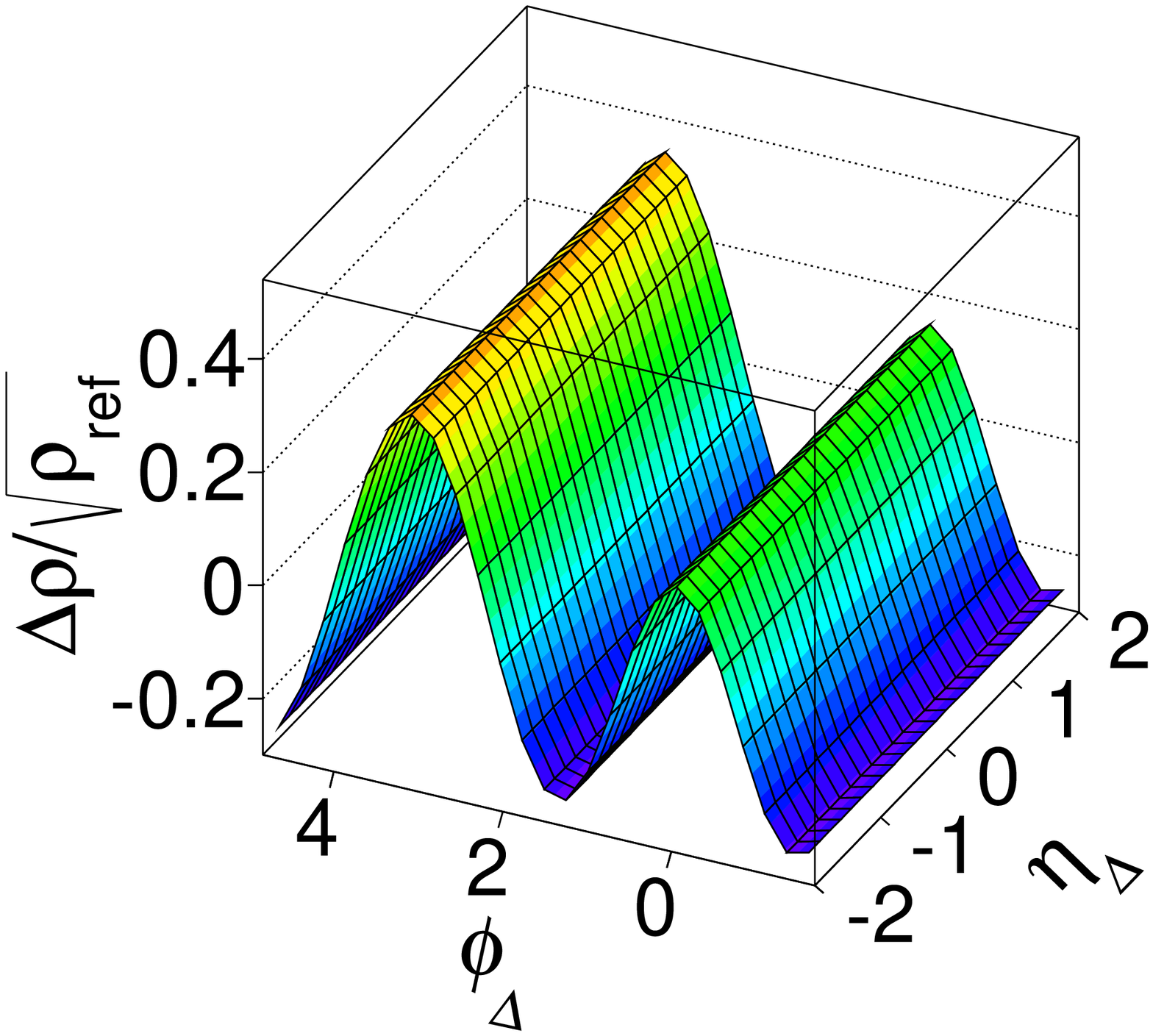}
\put(-40,75){\bf (b)}
\includegraphics[keepaspectratio,width=1.6in]{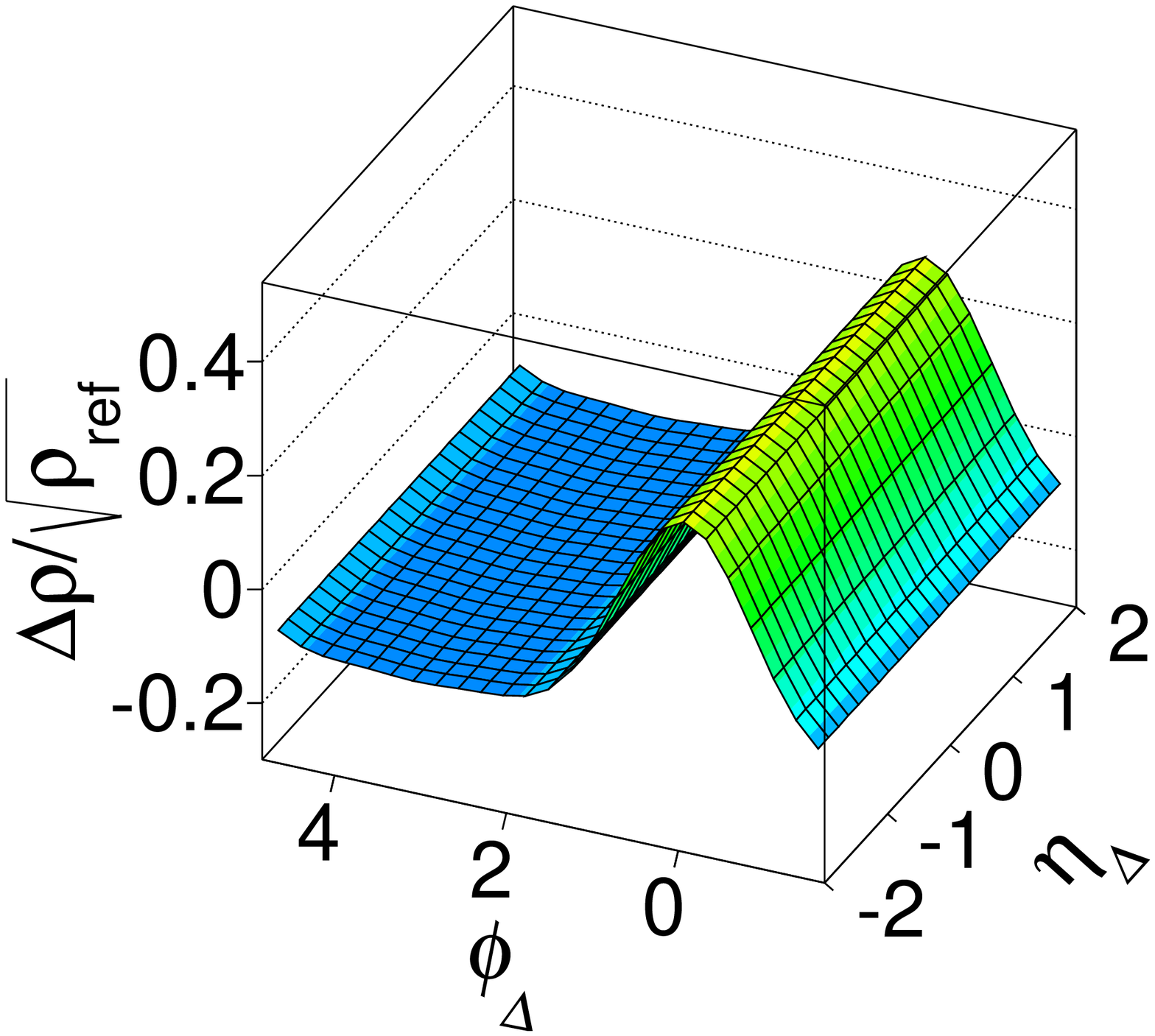}
\put(-50,77){\bf (c)}
\caption{\label{Fig1}
(Color online) Multipoles from fits to the 200 GeV Au+Au 9-18\% centrality data~\cite{axialCI} showing: (a) Fitted dipole + quadrupole. (b) Fitted dipole + quadrupole + sextupole. (c) Difference (b) - (a).}
\end{figure*}

The standard model function accurately describes the $\eta_\Delta$-independent structure in the away-side correlation data~\cite{aya,axialCI}. Including a sextupole forces the dipole and quadrupole terms to adjust to maintain a good fit. The net difference is shown in Fig.~\ref{Fig1} for model fits to more-central Au+Au correlation data. The right-most panel shows the quantity [$A_{\rm D}^{\prime}\cos(\phi_\Delta - \pi)/2 + 2A_{\rm Q}^{\prime}\cos(2\phi_\Delta) + 2A_{\rm S}^{\prime}\cos(3\phi_\Delta)$ $-$ $A_{\rm D}\cos(\phi_\Delta - \pi)/2 - 2A_{\rm Q}\cos(2\phi_\Delta)$] where primes indicate fitting parameters obtained with an included sextupole.  The net structural difference is a narrow, same-side 1D peak (effective ridge) on azimuth~\cite{axialCI,SextBash,Tomv3-1,Tomv3-2} which is accurately represented as a periodic 1D Gaussian. Fitting the data with an added sextupole element is statistically equivalent to fitting the data with an additional 1D same-side azimuth Gaussian (SSG) whose width is approximately 0.7.

\begin{figure*}
\centering
\includegraphics[keepaspectratio,width=3.6in]{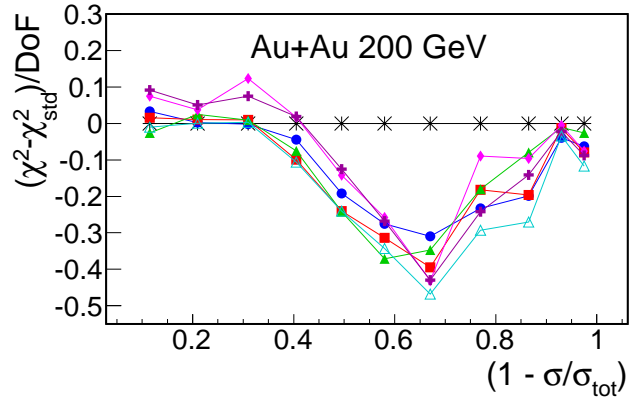}
\caption{\label{Fig2}
(Color online) Differences between the $\chi^2$/DoF for the six NG models and the standard model function versus centrality. Lines connect corresponding model results. The symbols denote each model as follows (see Ref.~\cite{SextBash}): (1) black stars, standard, (2) solid blue circles, sextupole, (3) solid red squares, SSG, (4) solid green triangles, NG exponents, (5) solid magenta diamonds, $\eta_\Delta$ polynomial, (6) solid purple ``plus'' symbols, $\eta_\Delta$ polynomial with NG $\phi_\Delta$ exponent, (7) open cyan triangles, quartic.}
\end{figure*}

The combination of an azimuth ridge and a 2D Gaussian produces a NG 2D peak. Projections of the same-side correlation data onto $\eta_\Delta$ are consistent with a 1D Gaussian within statistics. However, including small NG dependence improves the $\chi^2$~\cite{SextBash}. Two-dimensional NG fitting models~\cite{SextBash} are therefore considered further. NG modifications to the standard fitting model included: (i) replacing both exponents in the SS 2D Gaussian with fit parameters; (ii) replacing the $\eta_\Delta$-dependent Gaussian with a power series through terms of order $\eta_\Delta^4$; (iii) same as (ii) but allowing the exponent of the $\phi_\Delta$-dependent Gaussian to vary; (iv) adding quartic $\eta_\Delta^4$ and $\phi_\Delta^4$ terms in the argument of the exponential. The functional forms are given in~\cite{SextBash} . The sextupole term was excluded from fits which included these NG functions.

These five NG fitting models plus the standard model function with and without the sextupole were used to fit the angular correlation data for 200 GeV minimum-bias Au+Au collisions from STAR~\cite{axialCI}.
The best-fit values of $\chi^2$ per degree-of-freedom (DoF) for all models and collision centralities are plotted in Fig.~\ref{Fig2}. Centrality is represented by the fraction of total cross section $\sigma/\sigma_{\rm tot}$, where results for peripheral collisions are shown on the left-hand side.
From these results we find that all of the NG models reduce the $\chi^2$/DoF for the mid- to more-central collision data from $(1 - \sigma/\sigma_{\rm tot})$ = 0.4 to 0.9. The sextupole model is not special in that regard. For the NG models studied here the quartic model produces the best overall $\chi^2$/DoF.

Correlation measurements with higher $p_t$ cuts and for the higher collision energies attained at the LHC~\cite{atlas} provide strong evidence for NG dependence in the same-side 2D peak. It should not be surprising if a small NG dependence exists for same-side $p_t$-integral correlations at RHIC energies. Such occurrence would not exclude the possibility that the same-side correlation peak is dominated by pQCD jets with modified fragmentation~\cite{Tomjetfrag,Tommodfrag,BW}. While none of the NG fitting models considered here are excluded, it seems more plausible for possible NG structure in these data to originate locally in relative azimuth rather than arising from the combination of a same-side peaked structure with global angle correlations, such as $m > 2$ harmonics.

\section{Perturbative QCD models of the quadrupole correlation}

The scaling properties of the quadrupole correlation suggest that it may originate in the initial state via pQCD processes. Several authors recently presented pQCD based models in which a quadrupole correlation is generated by coherent gluon radiation from either BFKL Pomeron ladders~\cite{LR}, color dipoles~\cite{Boris}, or glasma~\cite{Raju}. Ref.~\cite{LR} provides explicit calculations which facilitate comparisons to data. Results from that paper are used here.

\begin{figure*}[t]
\centering
\includegraphics[keepaspectratio,width=2.3in]{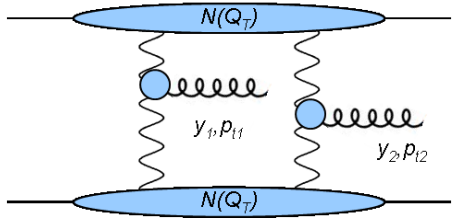}
\caption{\label{Fig3}
BFKL Pomeron diagrams with interfering gluon emission~\cite{LR}.}
\end{figure*}

In Ref.~\cite{LR} the BFKL-Pomeron diagram in Fig.~\ref{Fig3} results in quantum interference among the outgoing gluons such that the singles distribution contains a term proportional to $\cos2\phi$ where $\phi$ is measured relative to the N-N Pomeron momentum transfer $\vec{Q}_T$. The two-gluon density is
\bea
\label{Eq2}
\frac{d\sigma}{dy_1 dy_2 d^2p_{t1} d^2p_{t2}}
& = &
{\cal N} \left( 1 + \frac{1}{2} p^2_{t1} p^2_{t2} \langle \langle Q^4_T
\rangle \rangle \langle q^{-4} \rangle ^2 (2 + \cos2\phi_\Delta)
\right),
\eea
where ${\cal N}$ is proportional to the product of the single gluon distributions times the probability $N^2_{I\!\!Ph}(Q^2_T)$ of producing a two-Pomeron parton shower in a hadron-hadron collision. The momentum integrals were estimated in Ref.~\cite{LR} assuming a gluon saturation model with saturation scale $Q^2_S$, however, the unknown probability $N^2_{I\!\!Ph}(Q^2_T)$ was not estimated.

Each parton shower is assumed to produce a Poisson distribution with an average charged particle multiplicity $\bar{N}_{\rm ch}$ equal to the minimum-bias average multiplicity~\cite{LR} which is 2.5 per unit $\eta$ at midrapidity for $p+p$ at $\sqrt{s}$ = 200~GeV~\cite{Tompp}. The relative probability that each $p+p$ collision in a minimum-bias ensemble produces 1, 2, etc. parton showers is defined in this paper as $P_n$, $n \in [1,2,\cdots]$. In addition there is a finite probability of producing a hard-scattering process~\cite{Tompp} in each $p+p$ collision. These factors were combined in a model of the minimum-bias $p+p$ multiplicity frequency distribution, where the data are described with a negative binomial distribution (NBD). Fits to the latter provide estimates of $P_n$.  

The hard-scattering component of multiplicity in minimum-bias $p+p$ collisions at 200 GeV was studied in~\cite{Tompp}. Defining the soft and hard particle multiplicities as $n_{\rm s}$ and $n_{\rm h}$, where the total charged particle multiplicity $n_{\rm ch} = n_{\rm s} + n_{\rm h}$, it was found that
\bea
\label{Eq3}
n_{\rm h}/n_{\rm s} & = & \alpha n_{\rm ch}
\eea
where $\alpha = 0.005$ and 
\bea
\label{Eq3b}
n_{\rm h} & = & \alpha n_{\rm s}^2 / (1 - \alpha n_{\rm s}).
\eea
The frequency distribution on $n_{\rm s}$ in this model is
$\sum_{n=1} P_n {\cal P}(n_{\rm s},n\bar{N}_{\rm ch})$ where ${\cal P}(x,\bar{x})$ is the Poisson distribution on $x$ for mean $\bar{x}$. The hard component distribution depends on $n_{\rm s}$ and is proportional to ${\cal P}(n_{\rm h},\alpha n_{\rm s}^2 / (1 - \alpha n_{\rm s}))$. The joint probability distribution on $(n_{\rm h},n_{\rm s})$ projected onto total charge $n_{\rm ch}$ is
\bea
\label{Eq4}
\frac{1}{N_{\rm event}} \frac{dN_{\rm event}}{dn_{\rm ch}}
& = &
\sum_{n_{\rm s}} {\cal P}(n_{\rm ch} - n_{\rm s},\alpha n_{\rm s}^2 / (1 - \alpha n_{\rm s}))
\sum_{n=1} P_n {\cal P}(n_{\rm s},n\bar{N}_{\rm ch}).
\eea

\begin{figure*}[t]
\centering
\includegraphics[keepaspectratio,width=2.3in]{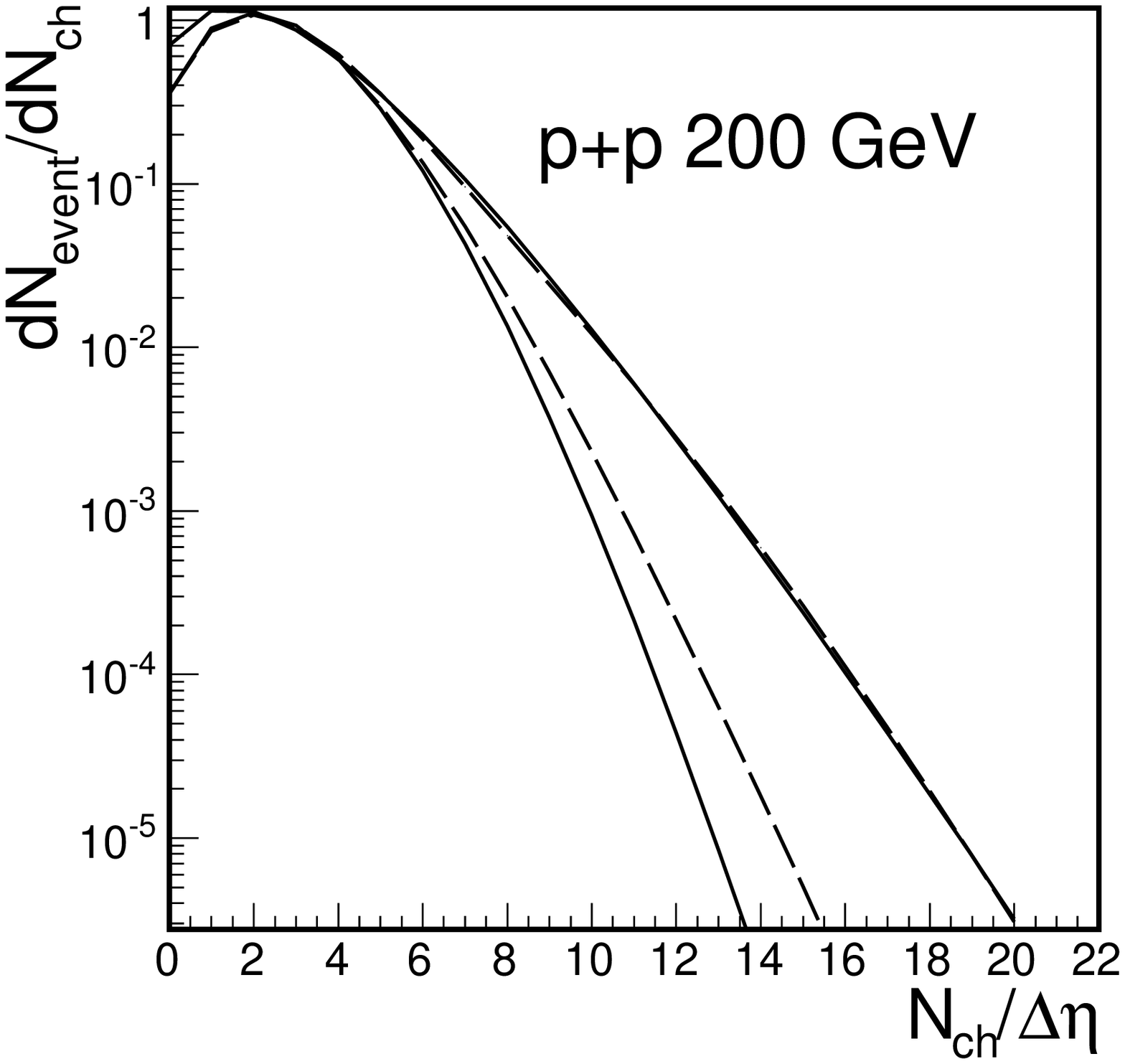}
\includegraphics[keepaspectratio,width=2.3in]{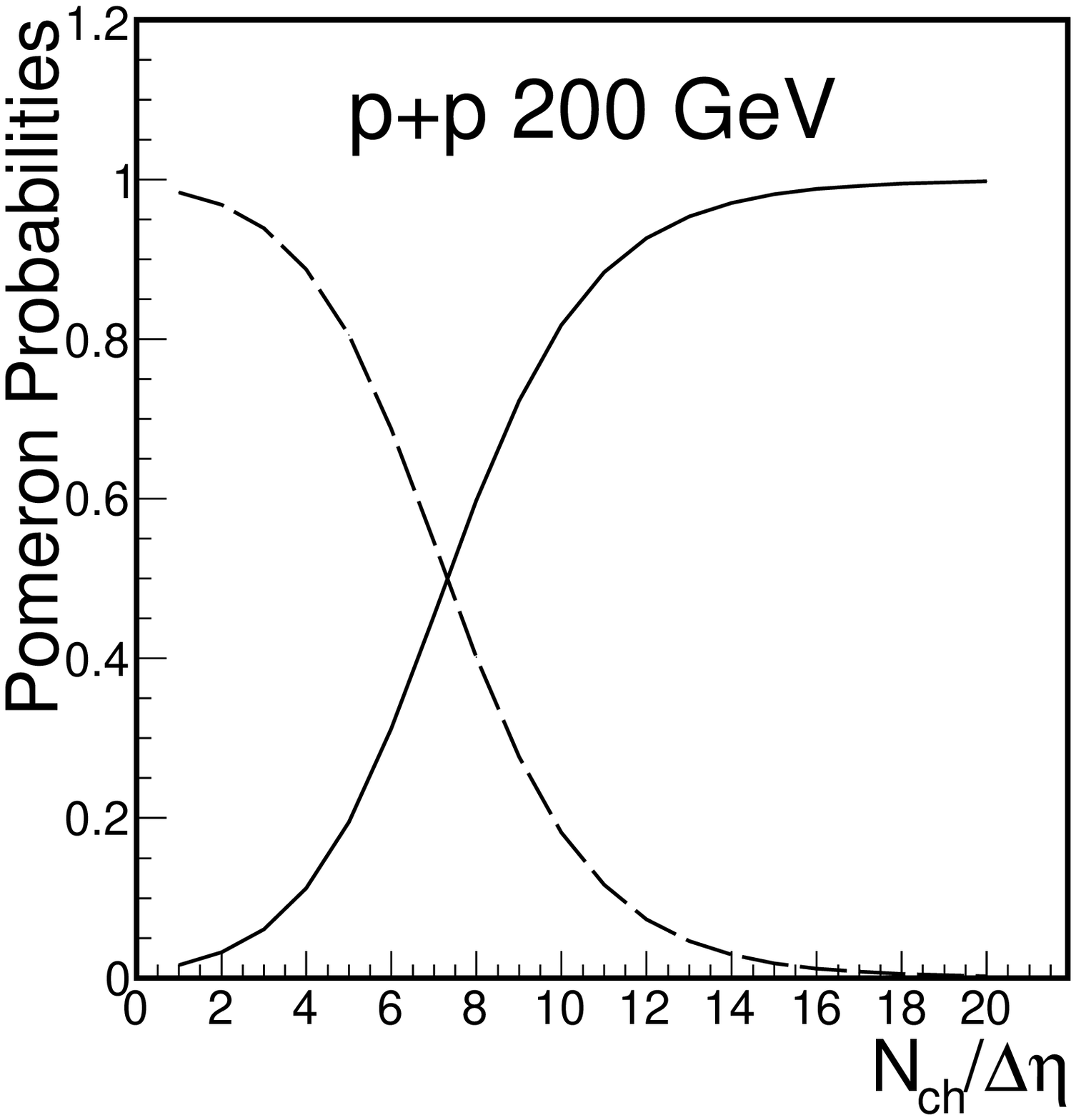}
\caption{\label{Fig4}
Left panel: One- and two-Pomeron shower distributions fitted to the 200 GeV $p+p$ minimum-bias frequency distribution for $|\eta|<0.5$. Right panel: Probabilities for one-Pomeron (dashed curve) and two-Pomeron (solid curve) showers.}
\end{figure*}

Fits to the data, shown in the left panel of Fig.~\ref{Fig4}, obtain $P_1 = 0.91$, $P_2 = 0.09$ and $P_{n>2} = 0$ for the minimum-bias average. The NBD representation of the data is shown by the upper solid curve. Distributions for one-Pomeron shower, one-Pomeron shower plus hard component, and the one- and two-Pomeron showers plus hard component fit are shown by the lower solid curve, lower dashed curve and upper dashed curve, respectively. The one-Pomeron (dashed) and two-Pomeron (solid) probabilities as a function of $n_{\rm ch}$ are shown in the right-hand panel of Fig.~\ref{Fig4}.  

The minimum-bias average quadrupole amplitude from Ref.~\cite{LR} is
\bea
\label{Eq5}
A_Q & = & \frac{\bar{N}_{\rm ch}}{2\pi\Delta\eta} \frac{P_2}{P_1 + 4P_2}
\langle p_t^2 \rangle^2 \langle \langle Q_T^4 \rangle \rangle
\langle q^{-4} \rangle^2
\eea
for $p+p$ collisions where only one- or two-Pomeron showers occur. Mean $p_t^2$ was estimated from spectrum data and equals 0.19~(GeV/$c$)$^2$. The momentum integrals were estimated in Ref.~\cite{LR} as $Q_S^{-4}$ in the fully saturated limit and as $m^4/(15Q_S^8)$ in the semi-saturated domain where $Q_S^2$ is assumed to be 0.6~(GeV/$c$)$^2$ and the dipole cut-off mass $m^2$ was assumed to be between 0.8 and 1.6~GeV$^2$. The quadrupole amplitude is predicted to be between 0.0003 and 0.003~\cite{correction} depending on the assumed gluon saturation model.  The measured 200 GeV $p+p$ minimum-bias quadrupole reported by the STAR experiment at this conference~\cite{Duncan} is 0.002 corresponding to azimuth asymmetry parameter $v_2 = 0.072$, a large value compared to typical $p_t$-integral $v_2$ values for Au+Au collisions. 

The $n_{\rm ch}$-dependent quadrupole amplitude is similarly predicted to be
\bea
\label{Eq6}
A_Q & = & P_2(n_{\rm ch}) \frac{n_{\rm ch} - 1}{8\pi\Delta\eta}
\langle p_t^2 \rangle^2 \langle \langle Q_T^4 \rangle \rangle
\langle q^{-4} \rangle^2.
\eea
The BFKL predicted quadrupole amplitude should increase with event multiplicity owing to the increasing 2-Pomeron probability shown in Fig.~\ref{Fig4}.  

Application of this model to proton + nucleus and nucleus + nucleus collisions can be done assuming a Glauber superposition approach.  The total number of correlated pairs in quantity $\Delta\rho$ for the quadrupole structure is an incoherent sum of those corresponding pairs from each nucleon + nucleon collision. If individual 2-Pomeron momentum transfer vectors $\vec{Q}_T$ are aligned via the initial overlap geometry of the colliding ions, then the total p+A and A+A quadrupole amplitudes will be further enhanced.

\section{Summary and Conclusions}

The physics implications of two-particle angular correlations from the RHIC and the LHC heavy-ion programs are intriguing. Most of the current interest concerns two major structures $-$ a jet related peak with its accompanying away-side dijet ridge, and a quadrupole.  Recent descriptions of the 2D angular correlation data, which are motivated by flow models, invoke higher harmonics ($m > 2$) to describe these data. In this work and in Ref.~\cite{SextBash} it was shown that the net effect of the $m > 2$ multipoles is to produce small, marginally significant NG dependence in the same-side peak's $\eta_\Delta$-dependent structure.  In my opinion the present results motivate a study of NG structure in the same-side 2D peak based on the fragmentation of minimum-bias jets in heavy-ion collisions.

The simultaneous appearance of approximately unperturbed minijets and large quad\-ru\-pole correlations combined with the latter's initial-state scaling properties suggest that an underlying pQCD mechanism may be responsible for the quadrupole correlation.  The BFKL Pomeron model of Levin and Rezaeian~\cite{LR} was shown to provide a predicted magnitude for 200 GeV minimum-bias $p+p$ collisions which is in reasonable agreement with recent STAR data. 

\section*{Acknowledgments}
  This work was supported in part by the U. S. Dept. of Energy grant No. DE-FG02-94ER40845.






\begin{thebibliography}{99}
\bibitem{correction}
The quadrupole values reported here are smaller than that presented during the conference due to an omitted factor of $1/2\pi$.
\end{thebibliography}
\end{document}